\documentclass[twocolumn,english,aps,prd, superscriptaddress,nofootinbib]{revtex4-2}
\usepackage[T1]{fontenc}
\usepackage[latin9]{inputenc}
\usepackage{color}
\usepackage{babel}
\usepackage{array}
\usepackage{booktabs}
\usepackage{units}
\usepackage{bm}
\usepackage{amsmath}
\usepackage{graphicx}
\usepackage[unicode=true,pdfusetitle,
 bookmarks=true,bookmarksnumbered=false,bookmarksopen=false,
 breaklinks=false,pdfborder={0 0 1},backref=false,colorlinks=true]
 {hyperref}

\makeatletter

\providecommand{\tabularnewline}{\\}

\renewcommand\toprule{\hline\hline}
\renewcommand\bottomrule{\hline\hline}

\hypersetup{citecolor=blue}

\makeatother

\begin{document}
\title{Invariant Jet Mass Measurements in $pp$ Collisions at $\sqrt{s}=\unit[200]{GeV}$
at RHIC}

\affiliation{Abilene Christian University, Abilene, Texas   79699}
\affiliation{AGH University of Science and Technology, FPACS, Cracow 30-059, Poland}
\affiliation{Alikhanov Institute for Theoretical and Experimental Physics NRC "Kurchatov Institute", Moscow 117218, Russia}
\affiliation{Argonne National Laboratory, Argonne, Illinois 60439}
\affiliation{American University of Cairo, New Cairo 11835, New Cairo, Egypt}
\affiliation{Brookhaven National Laboratory, Upton, New York 11973}
\affiliation{University of California, Berkeley, California 94720}
\affiliation{University of California, Davis, California 95616}
\affiliation{University of California, Los Angeles, California 90095}
\affiliation{University of California, Riverside, California 92521}
\affiliation{Central China Normal University, Wuhan, Hubei 430079 }
\affiliation{University of Illinois at Chicago, Chicago, Illinois 60607}
\affiliation{Creighton University, Omaha, Nebraska 68178}
\affiliation{Czech Technical University in Prague, FNSPE, Prague 115 19, Czech Republic}
\affiliation{Technische Universit\"at Darmstadt, Darmstadt 64289, Germany}
\affiliation{ELTE E\"otv\"os Lor\'and University, Budapest, Hungary H-1117}
\affiliation{Frankfurt Institute for Advanced Studies FIAS, Frankfurt 60438, Germany}
\affiliation{Fudan University, Shanghai, 200433 }
\affiliation{University of Heidelberg, Heidelberg 69120, Germany }
\affiliation{University of Houston, Houston, Texas 77204}
\affiliation{Huzhou University, Huzhou, Zhejiang  313000}
\affiliation{Indian Institute of Science Education and Research (IISER), Berhampur 760010 , India}
\affiliation{Indian Institute of Science Education and Research (IISER) Tirupati, Tirupati 517507, India}
\affiliation{Indian Institute Technology, Patna, Bihar 801106, India}
\affiliation{Indiana University, Bloomington, Indiana 47408}
\affiliation{Institute of Modern Physics, Chinese Academy of Sciences, Lanzhou, Gansu 730000 }
\affiliation{University of Jammu, Jammu 180001, India}
\affiliation{Joint Institute for Nuclear Research, Dubna 141 980, Russia}
\affiliation{Kent State University, Kent, Ohio 44242}
\affiliation{University of Kentucky, Lexington, Kentucky 40506-0055}
\affiliation{Lawrence Berkeley National Laboratory, Berkeley, California 94720}
\affiliation{Lehigh University, Bethlehem, Pennsylvania 18015}
\affiliation{Max-Planck-Institut f\"ur Physik, Munich 80805, Germany}
\affiliation{Michigan State University, East Lansing, Michigan 48824}
\affiliation{National Research Nuclear University MEPhI, Moscow 115409, Russia}
\affiliation{National Institute of Science Education and Research, HBNI, Jatni 752050, India}
\affiliation{National Cheng Kung University, Tainan 70101 }
\affiliation{Nuclear Physics Institute of the CAS, Rez 250 68, Czech Republic}
\affiliation{Ohio State University, Columbus, Ohio 43210}
\affiliation{Institute of Nuclear Physics PAN, Cracow 31-342, Poland}
\affiliation{Panjab University, Chandigarh 160014, India}
\affiliation{Pennsylvania State University, University Park, Pennsylvania 16802}
\affiliation{NRC "Kurchatov Institute", Institute of High Energy Physics, Protvino 142281, Russia}
\affiliation{Purdue University, West Lafayette, Indiana 47907}
\affiliation{Rice University, Houston, Texas 77251}
\affiliation{Rutgers University, Piscataway, New Jersey 08854}
\affiliation{Universidade de S\~ao Paulo, S\~ao Paulo, Brazil 05314-970}
\affiliation{University of Science and Technology of China, Hefei, Anhui 230026}
\affiliation{Shandong University, Qingdao, Shandong 266237}
\affiliation{Shanghai Institute of Applied Physics, Chinese Academy of Sciences, Shanghai 201800}
\affiliation{Southern Connecticut State University, New Haven, Connecticut 06515}
\affiliation{State University of New York, Stony Brook, New York 11794}
\affiliation{Instituto de Alta Investigaci\'on, Universidad de Tarapac\'a, Arica 1000000, Chile}
\affiliation{Temple University, Philadelphia, Pennsylvania 19122}
\affiliation{Texas A\&M University, College Station, Texas 77843}
\affiliation{University of Texas, Austin, Texas 78712}
\affiliation{Tsinghua University, Beijing 100084}
\affiliation{University of Tsukuba, Tsukuba, Ibaraki 305-8571, Japan}
\affiliation{United States Naval Academy, Annapolis, Maryland 21402}
\affiliation{Valparaiso University, Valparaiso, Indiana 46383}
\affiliation{Variable Energy Cyclotron Centre, Kolkata 700064, India}
\affiliation{Warsaw University of Technology, Warsaw 00-661, Poland}
\affiliation{Wayne State University, Detroit, Michigan 48201}
\affiliation{Yale University, New Haven, Connecticut 06520}

\author{M.~S.~Abdallah}\affiliation{American University of Cairo, New Cairo 11835, New Cairo, Egypt}
\author{J.~Adam}\affiliation{Brookhaven National Laboratory, Upton, New York 11973}
\author{L.~Adamczyk}\affiliation{AGH University of Science and Technology, FPACS, Cracow 30-059, Poland}
\author{J.~R.~Adams}\affiliation{Ohio State University, Columbus, Ohio 43210}
\author{J.~K.~Adkins}\affiliation{University of Kentucky, Lexington, Kentucky 40506-0055}
\author{G.~Agakishiev}\affiliation{Joint Institute for Nuclear Research, Dubna 141 980, Russia}
\author{I.~Aggarwal}\affiliation{Panjab University, Chandigarh 160014, India}
\author{M.~M.~Aggarwal}\affiliation{Panjab University, Chandigarh 160014, India}
\author{Z.~Ahammed}\affiliation{Variable Energy Cyclotron Centre, Kolkata 700064, India}
\author{I.~Alekseev}\affiliation{Alikhanov Institute for Theoretical and Experimental Physics NRC "Kurchatov Institute", Moscow 117218, Russia}\affiliation{National Research Nuclear University MEPhI, Moscow 115409, Russia}
% ...

\collaboration{STAR Collaboration}\noaffiliation

\begin{abstract}
We present the first inclusive measurements of the invariant and
SoftDrop jet mass in proton-proton collisions at $\sqrt{s}=\unit[200]{GeV}$
at STAR. The measurements are fully corrected for detector effects,
and reported differentially in both the jet transverse momentum and
jet radius parameter. We compare the measurements to established leading-order
Monte Carlo event generators and find that STAR-tuned PYTHIA-6 reproduces
the data, while LHC tunes of PYTHIA-8 and HERWIG-7 do not agree with
the data, providing further constraints on parameter tuning. Finally,
we observe that SoftDrop grooming, for which the contribution of wide-angle
non-perturbative radiation is suppressed, shifts the jet mass distributions
into closer agreement with the partonic jet mass as determined by
both PYTHIA-8 and a next-to-leading-logarithmic accuracy perturbative
QCD calculation. These measurements complement recent LHC measurements
in a different kinematic region, as well as establish a baseline for
future jet mass measurements in heavy-ion collisions at RHIC.
\end{abstract}
\maketitle

\section{Introduction\label{sec:Introduction}}

 A hard scattered parton will typically be highly virtual, and
therefore will quickly radiate a gluon \citep{JetsReview}. The cascade
of further radiation and splitting is called a parton shower, which
is described by the coupled differential DGLAP equations \citep{Dokshitzer:1977sg,Gribov:1972ri,Altarelli:1977zs}.
Once the partons' virtuality reach the non-perturbative (NP) regime,
they hadronize. In order to access the initial hard-scattered parton
and its evolution, the final-state particles are clustered into collective
objects called jets, using algorithms defined identically in theory
and experiment and robust to both arbitrarily soft and collinear radiation
\citep{antikt}. Jets have been used e.g., to refine the strong coupling
\citep[ Sec.  9.4.5]{PDG2020}, to search for new physics \citep{NewPhysicsUsingJets},
and to improve knowledge of the parton distribution functions (PDFs)
\citep{PDFsUsingJets}. Measurements of jet substructure specifically
test fundamental QCD via final state radiation patterns, and are important
in a heavy-ion context due to e.g., possible coherent energy loss
of hard scattered partons in the QCD medium \citep{CoherentRadiation}.
In this paper, we present the measurement of a jet substructure observable
called the jet mass, $M$, defined as the magnitude of the four-momentum
sum of the jet constituents, $M=\left|\sum_{i\in\text{jet}}p_{i}\right|=\sqrt{E^{2}-\bm{p}^{2}}$,
where $E$ and $\bm{p}$ are the energy and three-momentum of the
jet, respectively (see \citep{ATLAS1,ATLAS2,ATLAS3,CDF,CMS1,CMS2,CMS3,ALICE}
for previous measurements at the LHC and the Tevatron). The influence
of the initial hard scattering, fragmentation, and hadronization on
the resulting jet angular and momentum scales implies that the jet
mass is sensitive to the details of these processes. The mass of jets
arising from heavy quarks is also sensitive to the initiating quark's
mass (see Ref.~\citep{CMS2} for an extraction of the top quark mass
from top quark jets) although in the kinematic regime of this paper,
light quark jets are dominant \citep{STARtune}, and access instead
the initiating parton's virtuality. We can use the jet mass distributions
to test the applicability of perturbative QCD (pQCD) calculations
at low jet energies and also to tune parton shower parameters in Monte
Carlo (MC) event generators such as PYTHIA and HERWIG \citep{P6,P8,H7}
for better prediction power across varying center-of-mass energies.

Jet mass is sensitive to both perturbative and non-perturbative physics.
At the Relativistic Heavy Ion Collider (RHIC), the lower center-of-mass
energy, $\sqrt{s}$,  compared to the Large Hadron Collider (LHC),
reduces the importance of higher-order (N$^{n}$LO) effects, while
the lower jet transverse momentum, $p_{\text{T},\text{jet}}$, increases
the sensitivity to the underlying event (UE) and hadronization \citep{UE}.
However, one may reduce contributions of NP physics such as UE and
hadronization in a Sudakov-safe  way \citep{Sudakov} with the SoftDrop
grooming algorithm \citep{SoftDrop} which is described in more detail
in Sec.~\ref{sec:Experimental-setup}. Therefore, a groomed jet
mass measurement, i.e., a jet mass measurement on a population of
jets to which the SoftDrop grooming algorithm has been applied, allows
a more direct comparison with analytic calculations \citep{GroomingTheoryPaper}
at varying orders. We would also expect that the Sudakov peak --
the result of a small probability of low and high mass jets due to
suppression of perfectly collinear gluon radiation after resummation
and hard or wide-angle radiation, respectively -- shifts to lower
jet mass after SoftDrop grooming due to the removal of wide-angle
radiation  \citep{SudakovPeak}. Additionally, a comparison of groomed
jet mass to ungroomed jet mass differentially in $p_{\text{T,jet}}$,
and the jet radius parameter, $R$, can be utilized to assess experimentally
the phase space for which the magnitude of NP effects is significant.

The jet mass is also a useful quantity in heavy-ion collisions (see
\citep{ALICE} for a previous measurement at the LHC), in which any
hard scatters occur before the hot, dense, colored QCD medium is formed.
The partons, which also carry color charge, then interact with the
medium which may temporarily increase the virtuality of the jets causing
an increase in gluon radiation (increasing the jet mass), while on
the whole, virtuality will decrease more quickly in the medium (decreasing
the jet mass) \citep{ALICE}. Therefore, this measurement will serve
as a vacuum baseline for a future measurement in heavy-ion collisions,
in which medium modification of the jet mass in a hot, dense nuclear
environment may be observed \citep{JoernAbhijit}.

In $pp$ collisions, there have been several recent jet mass measurements
\citep{ATLAS1,ATLAS2,ATLAS3,CDF,CMS1,CMS2,CMS3} (almost exclusively
at the LHC) and calculations \citep{Theory1,Theory2,Theory3,NGLsJetMass,Theory4,Theory5,Theory6,Theory7}
(at LHC kinematics) demonstrating that LHC-tuned MCs and calculations
are able to describe the data at LHC energies. This paper presents
the first fully corrected inclusive jet mass measurement at RHIC energies.

In this paper, we present the ungroomed and groomed jet mass differentially
in both $p_{\text{T,jet}}$ and $R$, where jets are reconstructed
with the FastJet anti-$k_{\text{T}}$ algorithm \citep{FastJet,antikt}.
We compare to three MC simulations in this analysis: PYTHIA-6, HERWIG-7,
and PYTHIA-8. We also show a PYTHIA-8 parton-level simulation and
a pQCD calculation at next-to-leading-logarithmic (NLL) accuracy to
emphasize the suppression of NP effects by grooming.

\section{Experimental setup\label{sec:Experimental-setup}}

The data used for this analysis were collected by the Solenoidal Tracker
at RHIC (STAR) detector in $pp$ collisions at $\sqrt{s}=\unit[200]{GeV}$
in 2012. Charged tracks are reconstructed via the Time Projection
Chamber (TPC) \citep{TPC}, and the surrounding Barrel Electromagnetic
Calorimeter (BEMC) \citep{BEMC} measures electromagnetic energy deposits
in its 4800 towers each covering $0.05\times0.05$ in pseudorapidity
($\eta$) and azimuth ($\phi$). These detectors have full azimuthal
coverage and $\left|\eta\right|<1$. In a procedure called hadronic
correction, the tower energy is corrected for energy from tracks measured
in the TPC which match to the tower, to avoid double-counting. Any
negative corrected tower energies are set to zero. This procedure
is optimal with respect to the result on the jet momentum resolution
and neutral energy fraction \citep{HadronicCorrection}. The BEMC
is additionally used online as an event trigger that requires a total
ADC value above a certain threshold, corresponding to $\sum E_{\text{T},\text{tower}}>\unit[7.3]{GeV}$,
in one of 18 partially overlapping $1.0\times1.0$ $\eta\times\phi$
groupings of towers called Jet Patches \citep{RaghavArXiv}.

For the analysis we impose certain quality requirements on the tracks
and towers, e.g., $0.2<p_{\text{T}}<\unit[30]{GeV}/c$, $0.2<E_{\text{T}}<\unit[30]{GeV}$,
and $\left|\eta\right|<1$, as well as standard overall event quality
assurance cuts \citep{RaghavArXiv}. Namely, we require the $z$-component
of the primary vertex location to be within $\unit[30]{cm}$ of the
center of the detector, and due to worsening momentum resolution for
high-$p_{\text{T}}$ tracks, we reject events containing tracks with
$p_{\text{T}}>\unit[30]{GeV}/c$ or towers with $E_{\text{T}}>\unit[30]{GeV}$.
Although three of the components of the particle or tower four-vector
are specified, the mass is unknown, so we choose a mass assignment.
This analysis does not attempt any particle identification, so at
detector-level (i.e.,~data and simulation in which particles traverse
detector material), the pion mass is assumed for charged tracks, and
tower energy deposits are assumed massless; at particle-level (i.e.,~vacuum
simulation), the particle PDG mass \citep{PDG2020} is assigned (as
in Ref.~\citep{ALICE}). The effect of the difference between detector-level
and particle-level caused by this choice of mass assignments is folded
into the detector response, and is corrected via unfolding so that
the reported particle-level data is comparable to MC simulation and
analytic calculations using the hadrons' PDG mass.

We cluster accepted tracks and towers into jets using the anti-$k_{\text{T}}$
sequential recombination algorithm with the $E$-scheme in the FastJet
framework. This algorithm defines distance metrics 
\[
d_{ij}=\min\left(p_{\text{T},i}^{-2},p_{\text{T},j}^{-2}\right)\dfrac{\Delta_{ij}^{2}}{R^{2}}
\]
and $d_{iB}=p_{\text{T},i}^{-2}$, where $\Delta_{ij}^{2}=\left(y_{i}-y_{j}\right)^{2}+\left(\phi_{i}-\phi_{j}\right)^{2}$,
and $y_{i}$ is the rapidity of object $i$. It then finds the smallest
distance among objects; it recombines $i$ and $j$ if $d_{ij}$ is
the smallest distance, or removes $i$ from consideration and considers
it a jet if $d_{iB}$ is the smallest distance. This is repeated until
only jets remain. In the analysis, jets are selected to be contained
within the experimental fiducial volume of the TPC and BEMC ($\left|\eta_{\text{jet}}\right|<1-R$,
where $\eta_{\text{jet}}$ is the pseudorapidity of the jet axis).
In addition, we consider only jets with less than $90\%$ of their
energy from the BEMC to reduce the contribution of beam background
\citep{RaghavArXiv}. After all event, track, tower, and jet cuts,
the remaining jet population is considered for the ungroomed mass
analysis.

The SoftDrop declustering algorithm is then applied to these jets
to obtain a population of groomed jets. That is, jets are reclustered
in an angular-ordered tree (with the Cambridge-Aachen algorithm \citep{CAcambridge,CAaachen}),
and starting with the outermost constituent pairs (denoted with the
subscripts $1$ and $2$), those failing the condition 
\[
\dfrac{\min\left(p_{\text{T},1},p_{\text{T},2}\right)}{p_{\text{T},1}+p_{\text{T},2}}>z_{\text{cut}}\left(\dfrac{\Delta_{12}}{R}\right)^{\beta}
\]
have the softer constituent removed; the procedure is iterated until
either a pair passes the condition (``grooming mode'') or the jet
can no longer be declustered (``tagging mode''). Here, $\beta$
controls the extent to which wide-angle constituents are removed,
and $z_{\text{cut}}$ is the momentum sharing fraction threshold for
the pair. E.g., $z_{\text{cut}}=0.2$, $\beta=0$ denotes that the
softer constituent must carry 20\% of the overall momentum of the
pair, with no angular consideration. We apply the SoftDrop grooming
procedure in tagging mode with $z_{\text{cut}}=0.1$ and $\beta=0$
to these ungroomed jets, to obtain the population of jets to be considered
for the groomed mass analysis. This canonical choice of parameters
simplifies calculations \citep{SoftDrop}. It was also shown, ibid.,
in a PYTHIA-8 study at LHC energies that an observable similar to
jet mass has very little contribution from hadronization corrections
or underlying event, after grooming with the parameters used in this
analysis. This provides evidence for the suppression of NP effects
by the SoftDrop declustering procedure. In this paper, observables
on the set of groomed jets are subscripted with a ``g'', and are
shown for ranges of the corresponding ungroomed $p_{\text{T},\text{jet}}$,
to allow a direct comparison between jet mass and groomed jet mass.

\section{Correction for detector effects\label{sec:Unfolding}}

\begin{figure}
\begin{centering}
\includegraphics[width=1\columnwidth]{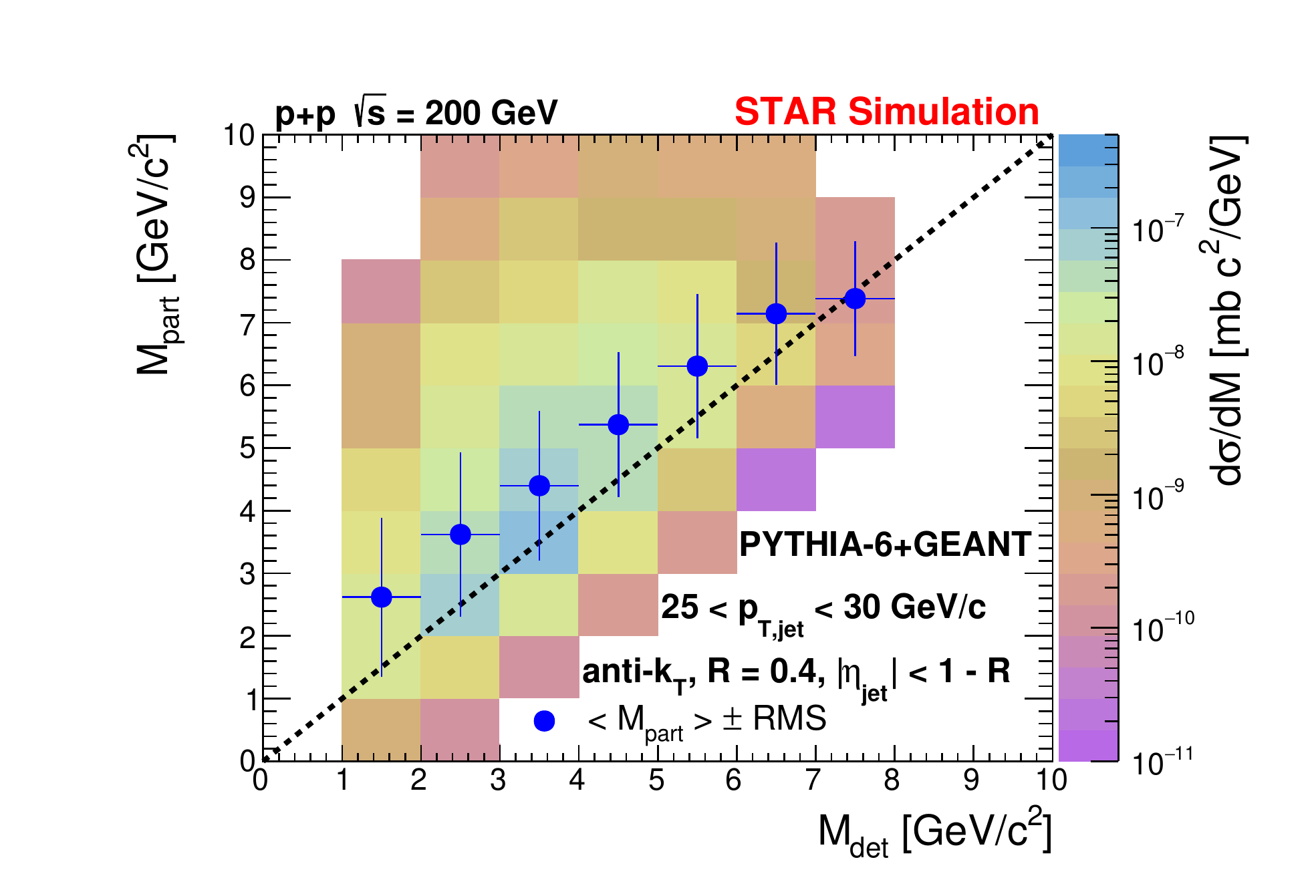}
\par\end{centering}
\caption{Correlation between simulated jet mass at particle-level (given by
PYTHIA-6) and detector-level (given by PYTHIA-6+GEANT) for a single
$p_{\text{T,jet}}$ selection at both particle- and detector-levels
($25<p_{\text{T,jet}}<\unit[30]{GeV}/c$) and jet radius parameter
$R=0.4$. The mean particle-level mass for a particular detector-level
mass selection is given by the blue circles, with horizontal bars
spanning the detector-level jet mass bin width and vertical bars denoting
the particle-level jet mass RMS for that selection of $M_{\text{det}}$.\label{fig:resies}}
\end{figure}
To correct for detector effects, we use a two-dimensional Bayesian
unfolding procedure \citep{Bayes}; with this we obtain the physical
distributions (``causes'') most likely to have lead to the observed
data (``effects''), using Bayesian inference. We construct two detector
response matrices, one for $p_{\text{T,jet}}$ and $M$, and one for
$p_{\text{T,jet}}$\footnote{Note that throughout, $M_{\text{g}}$ is shown as a function of $p_{\text{T}}$
of the ungroomed (rather than groomed) jet.} and $M_{\text{g}}$, and unfold (in $M$ and $M_{\text{g}}$ separately)
the uncorrected data to particle-level via the RooUnfold package \citep{RooUnfold}
with a regularization parameter of 4. This parameter corresponds to
the number of iterations of the Bayesian inference procedure: each
subsequent iteration uses as a prior the posterior from the previous
iteration, so that larger numbers of iterations yield less reliance
on the initial prior distribution given in this case by Monte Carlo.
However, beyond some number of iterations, the statistical uncertainties
grow rapidly, hence the choice of 4.

\begin{figure*}
\begin{centering}
\includegraphics[width=1\textwidth]{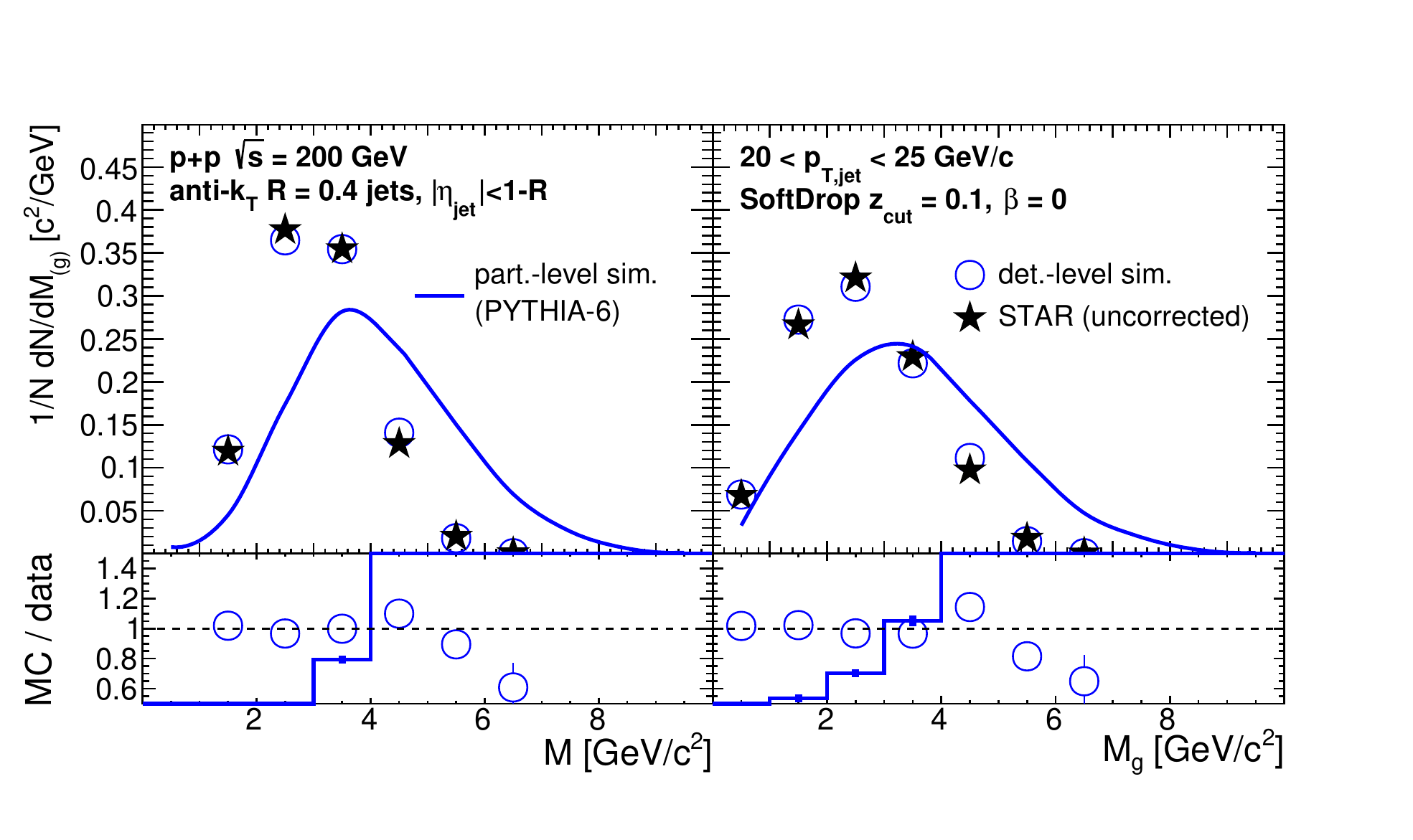}
\par\end{centering}
\caption{Comparison of STAR data (full star markers), particle-level simulation
given by PYTHIA-6 (Perugia 2012 STAR Tune, solid blue line), and detector-level
simulation given by PYTHIA-6+GEANT (open blue circles) for jet mass,
$M$ (left) and groomed jet mass, $M_{\text{g}}$ (right) for a single
$p_{\text{T,jet}}$ selection ($20<p_{\text{T,jet}}<\unit[25]{GeV}/c$)
and $R=0.4$. Data are uncorrected for detector effects. Statistical
uncertainties are smaller than the size of the markers. We quantitatively
compare via ratio in the lower panels to the particle- and detector-level
simulations.\label{fig:rawmass}}
\end{figure*}

A visualization of a subset of the four-dimensional response matrix
is shown in Fig.~\ref{fig:resies}, with particle-level (detector-level)
jet mass denoted $M_{\text{part(det)}}$. To construct the matrix,
we begin with a sample of $pp$ events at $\sqrt{s}=\unit[200]{GeV}$
generated by the PYTHIA-6.428 event generator with CTEQ6L1 PDFs \citep{CTEQ6PDFs}
and the Perugia 2012 tune, further tuned to STAR data (see Refs.~\citep{STARtune,RaghavArXiv}
for details). These particle-level events are then passed to the
GEANT-3 \citep{GEANT3} STAR detector simulation to obtain the detector
hits, and are combined with zero-bias (randomly triggered) data from
the same $pp$ run period. Detector-level events, from the GEANT simulation,
are treated exactly like events in the data as described in Sec.~\ref{sec:Experimental-setup}.
The (groomed) jet mass from PYTHIA-6 and PYTHIA-6+GEANT is shown on
the (right) left in Fig.~\ref{fig:rawmass} with a comparison to
the uncorrected data, for $R=0.4$ jets with $20<p_{\text{T,jet}}<\unit[25]{GeV}/c$.
We observe good agreement between the uncorrected data and detector
simulation, as shown in the ratio panel. Similar agreement is found
in all three of the $p_{\text{T,jet}}$ selections reported in this
paper. Note that the ungroomed detector-level jet mass is required
to be greater than $\unit[1]{GeV}/c^{2}$ to improve the performance
of the unfolding procedure. The groomed mass has no such restriction.

\begin{figure*}
\begin{centering}
\includegraphics[width=1\textwidth]{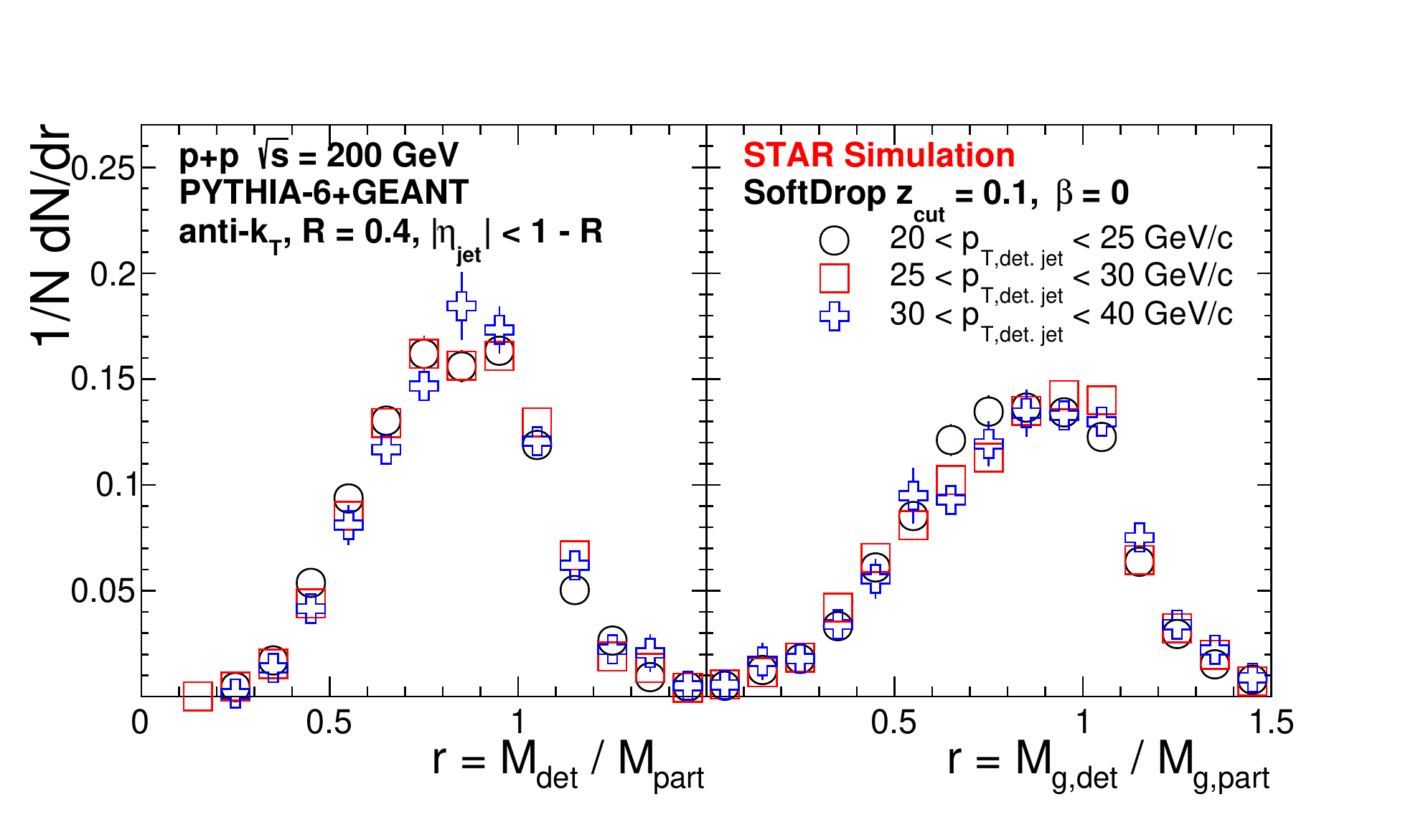}
\par\end{centering}
\caption{Ratio of simulated detector-level to particle-level jet mass (ungroomed
and groomed, respectively) for three selections of detector-level
$p_{\text{T,jet}}$ (with no selection on the corresponding particle-level
$p_{\text{T,jet}}$), for $R=0.4$ particle-level jets from PYTHIA-6
matched event-by-event to detector-level jets from PYTHIA-6+GEANT.
Most statistical uncertainties -- denoted by vertical bars -- are
smaller than the size of the markers.\label{fig:response}}
\end{figure*}

To compare the particle- and detector-level jets on a jet-by-jet basis,
we match geometrically by requiring potential matches to pass the
criterion,
\[
\sqrt{\left(\Delta\eta\right)^{2}+\left(\Delta\phi\right)^{2}}<R,
\]
where $\Delta\eta$ ($\Delta\phi$) denotes the difference between
the particle- and detector-level jet axes in $\eta$ ($\phi$). A
match is a jet pair that satisfies the above criterion and falls in
the jet momentum and jet mass ranges of the response matrix (namely
at particle-level, $p_{\text{T,jet}}\in\unit[\left(5,80\right)]{GeV}/c$,
and at detector-level $p_{\text{T,jet}}\in\unit[\left(15,60\right)]{GeV}/c$
and $M>\unit[1]{GeV}/c^{2}$). The detector performance is quantified
by a ratio, $r$, of detector-level to particle-level mass on a jet-by-jet
basis for the matched pairs. This is shown in Fig.~\ref{fig:response}
for jet mass on the left and groomed mass on the right, for $R=0.4$
jets with $20<p_{\text{T,jet}}<\unit[25]{GeV}/c$ as black circles,
$25<p_{\text{T,jet}}<\unit[30]{GeV}/c$ as red squares, and $30<p_{\text{T,jet}}<\unit[40]{GeV}/c$
as blue crosses. Tracking inefficiency of the detector reduces the
jet mass, shifting the peak of the distribution to the left of unity.
We observe that the resolution (the width of the distribution) is
independent of $p_{\text{T},\text{jet}}$, which is beneficial for
numerical stability in the unfolding. There are relatively more groomed
jets than ungroomed jets with low $r$ (say, $r<0.5$), due to the
lack of a minimum detector-level $M_{\text{g}}$ requirement mentioned
above. We use the resolution to determine the appropriate jet mass
bin width ($\unit[1]{GeV}/c^{2}$) used in the unfolding procedure,
while the $p_{\text{T,jet}}$ bin width ($\unit[5]{GeV}/c$) is the
same as what was used in Ref.~\citep{RaghavArXiv}. The highest $p_{\text{T,jet}}$
selection for which we show the jet mass ($30<p_{\text{T,jet}}<\unit[40]{GeV}/c$)
is wider, as in that paper, in order to improve the low statistics
at higher $p_{\text{T,jet}}$ due to the steeply falling $p_{\text{T,jet}}$
spectrum at RHIC.

\section{Systematic uncertainties}

There are four sources of systematic uncertainty involved in the jet
mass analysis. Two detector uncertainties were considered: uncertainty
on the tower gain calibration (3.8\%) and on the tracking efficiency
(4\%) \citep{STARtune}. Two procedural uncertainties were considered
as well. The first, the hadronic correction, is varied from the nominal
subtraction of 100\% of a track's $p_{\text{T}}$ from the matched
tower, to 50\% which is between the average charged hadron energy
deposition in the BEMC and the nominal over-subtraction of 100\%.
 In addition, there is uncertainty arising from the unfolding procedure
\citep{RaghavArXiv}. 

The unfolding procedure outlined in Sec.~\ref{sec:Unfolding} is
a dominant source of systematic uncertainties. We vary the regularization
parameter from the nominal value of 4 down to 2 (the minimum for an
iterative Bayesian approach) and up to 6 (above which the influence
of statistical fluctuations is large). We also vary the shape of the
particle-level $p_{\text{T,jet}}$ spectrum prior distribution given
by PYTHIA-6 as well as the detector-level $p_{\text{T,jet}}$ spectrum
given by PYTHIA-6+GEANT. Similarly, we vary the shape of the particle-level
mass spectrum prior distribution given by PYTHIA-6, which is adjusted
by the ratios of PYTHIA-6 to PYTHIA-8 and to HERWIG-7. The two particle-level
Bayesian prior smearings contribute roughly equally to the relative
systematic uncertainty, while the detector-level $p_{\text{T,jet}}$
spectrum variation is subdominant. The maximum envelope of these variations
is taken as the unfolding uncertainty.

The individual sources and total systematic uncertainties (the quadrature
sum of individual contributions) are shown in Table~\ref{tab:systs}
for representative $p_{{\rm T,jet}}$ and $M_{{\rm (g)}}$ selections.
Relative uncertainty values in Table~\ref{tab:systs} were obtained
by propagating each variation through the unfolding procedure, and
dividing the resulting fully corrected mass by the nominal result.
\begin{table*}
\begin{centering}
\begin{tabular}{>{\centering}m{2.1cm}>{\centering}m{1.65cm}>{\centering}m{0.9cm}>{\centering}m{1.6cm}>{\centering}m{1.6cm}>{\centering}m{1.9cm}}
\toprule 
Source / \linebreak Range in $M$ & Hadronic \linebreak Correction & Tower \linebreak Gain & Tracking \linebreak Efficiency & Unfolding \linebreak Procedure & Total \linebreak Systematics\tabularnewline
\midrule
$\left(1,2\right)$ GeV/$c^{2}$ & 1.3\% & 0.9\% & 13.0\% & 12.2\% & 17.9\%\tabularnewline
$\left(4,5\right)$ GeV/$c^{2}$ & 0.1\% & 0.6\% & 0.4\% & 4.1\% & 4.1\%\tabularnewline
$\left(7,8\right)$ GeV/$c^{2}$ & 3.6\% & 0.4\% & 6.9\% & 22.9\% & 24.1\%\tabularnewline
\bottomrule
\end{tabular}
\par\end{centering}
\medskip{}

\begin{centering}
\begin{tabular}{>{\centering}m{2.1cm}>{\centering}m{1.65cm}>{\centering}m{0.9cm}>{\centering}m{1.6cm}>{\centering}m{1.6cm}>{\centering}m{1.9cm}}
\toprule 
Source / \linebreak Range in $M_{\text{g}}$ & Hadronic \linebreak Correction & Tower \linebreak Gain & Tracking \linebreak Efficiency & Unfolding \linebreak Procedure & Total \linebreak Systematics\tabularnewline
\midrule
$\left(1,2\right)$ GeV/$c^{2}$ & 2.6\% & 0.7\% & 6.8\% & 9.1\% & 11.7\%\tabularnewline
$\left(4,5\right)$ GeV/$c^{2}$ & 1.0\% & 0.8\% & 1.5\% & 3.4\% & 4.0\%\tabularnewline
$\left(7,8\right)$ GeV/$c^{2}$ & 1.1\% & 0.2\% & 8.0\% & 28.3\% & 29.4\%\tabularnewline
\bottomrule
\end{tabular}
\par\end{centering}
\caption{Systematic uncertainties for an example jet population with $25<p_{\text{T,jet}}<\unit[30]{GeV/}c$
and, from the top to the bottom row, $1<M_{\text{(g)}}<2$, $4<M_{\text{(g)}}<5$,
and $7<M_{\text{(g)}}<\unit[8]{GeV}/c^{2}$. The total systematic
uncertainties are obtained by adding the sources in the four preceding
columns in quadrature. Upper: ungroomed jet mass. Lower: groomed jet
mass. \label{tab:systs}}
\end{table*}

\section{Results}

\begin{figure*}
\begin{centering}
\includegraphics[width=0.95\textwidth]{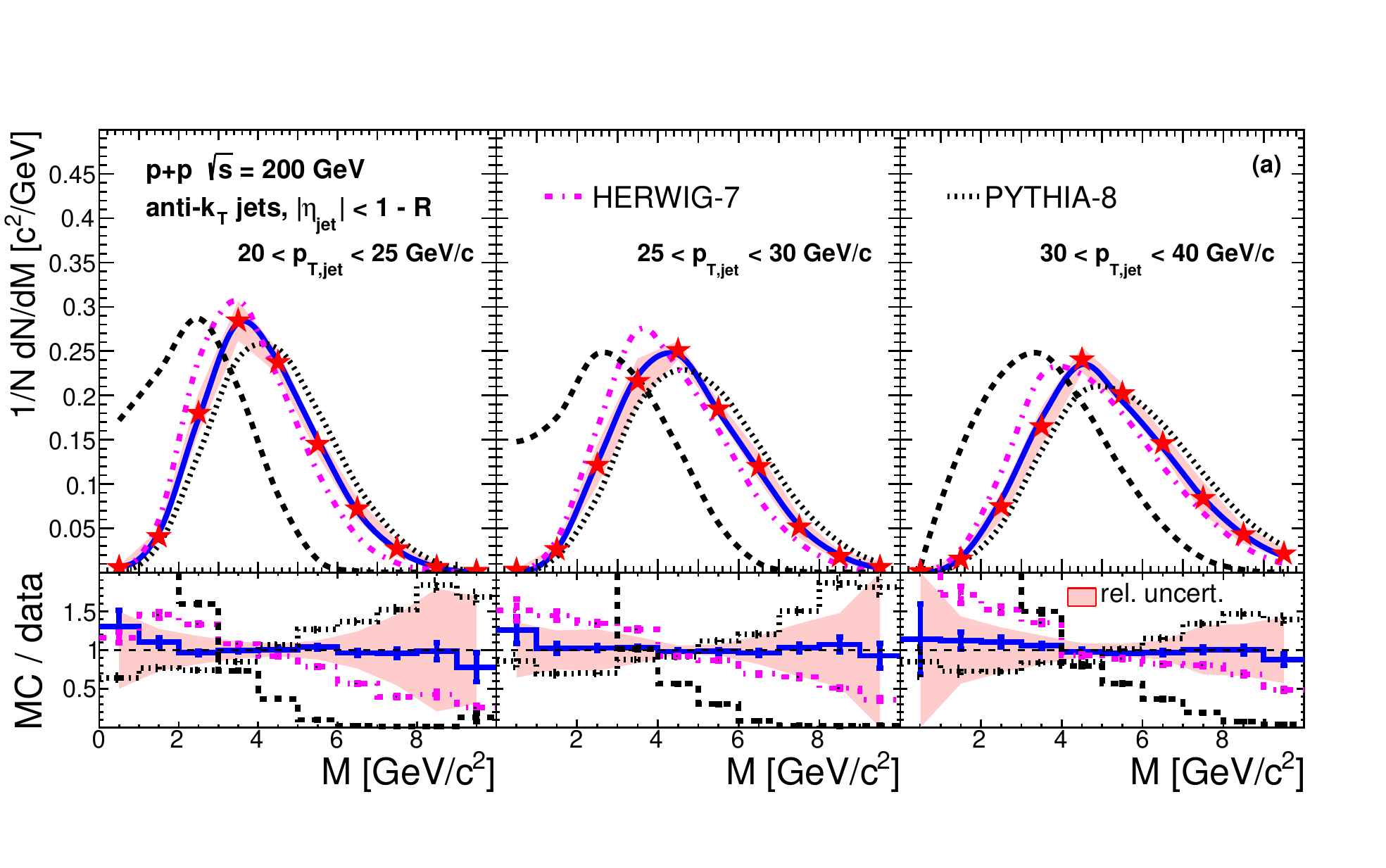}
\par\end{centering}
\begin{centering}
\vspace{-27bp}
\includegraphics[width=0.95\textwidth]{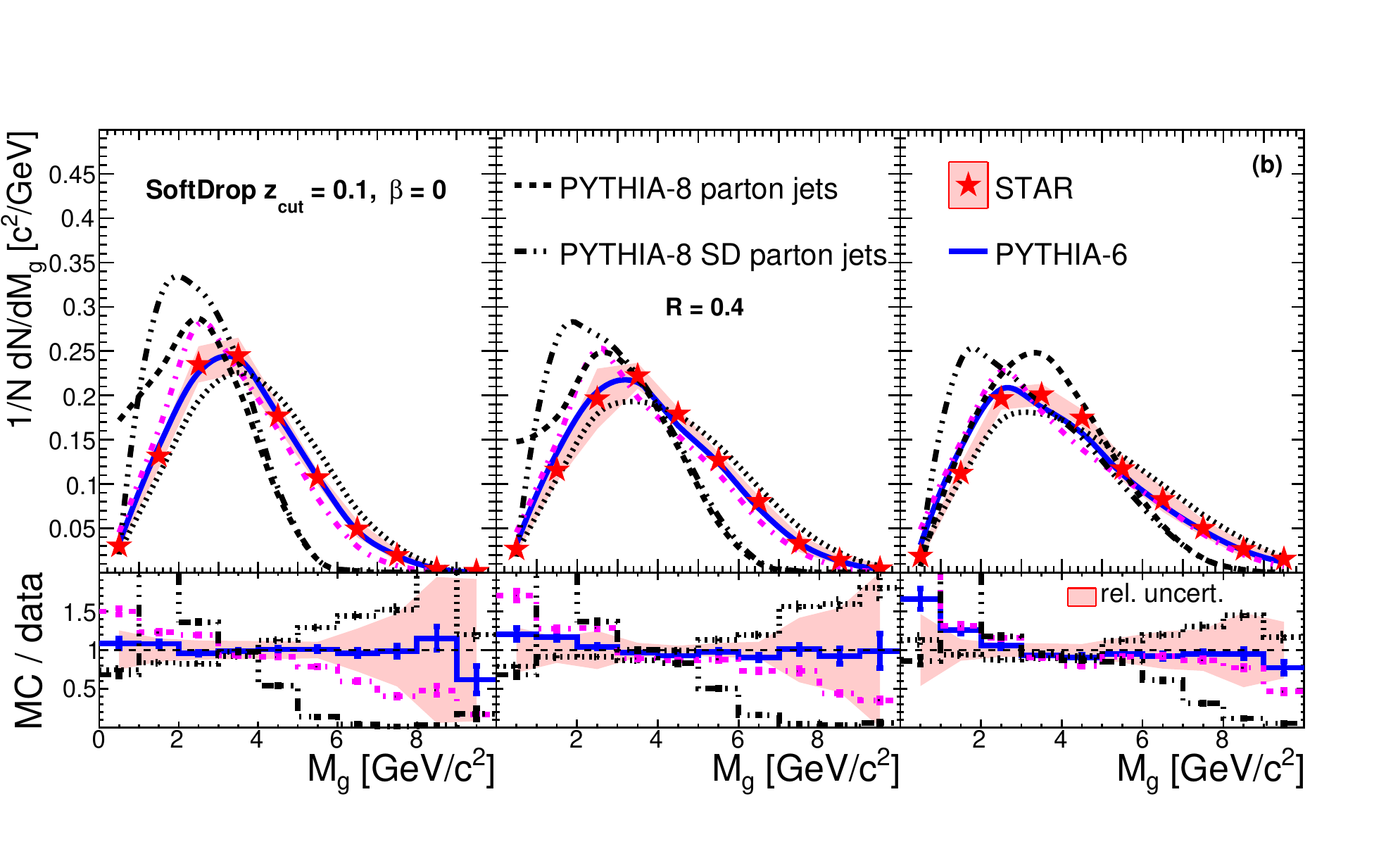}
\par\end{centering}
\caption{Distributions of (a) $M$ and (b) $M_{\text{g}}$ for $R=0.4$ anti-$k_{\text{T}}$
jets in $pp$ collisions at $\sqrt{s}=\unit[200]{GeV}$. From left
to right, the $p_{\text{T,jet}}$ increases from $20<p_{\text{T},\text{jet}}<\unit[25]{GeV}/c$
to $30<p_{\text{T},\text{jet}}<\unit[40]{GeV}/c$. The fully corrected
data (with accompanying red shaded band denoting systematic uncertainties)
are shown as solid red star markers. We quantitatively compare via
ratio (\textquotedblleft MC$/$data\textquotedblright ) to PYTHIA-6
(Perugia 2012 STAR Tune, solid blue line), PYTHIA-8 (Monash 2013 Tune,
dotted black line), and HERWIG-7 (EE4C Tune, dot-dashed magenta line)
in the lower part of each panel, along with a total relative systematic
uncertainty (\textquotedblleft rel. uncert.\textquotedblright ) on
the data (corresponding to the absolute systematic uncertainty given
by the shaded red band in the upper part of the panels). PYTHIA-8
with hadronization turned off is also shown as \textquotedblleft PYTHIA-8
parton jets\textquotedblright{} (dashed black line). Statistical uncertainties
are smaller than the size of the markers. In (b), we additionally
compare to PYTHIA-8 parton-level $M$ (dashed black line) and $M_{\text{g}}$
(dot-dashed black line, denoted \textquotedblleft SD\textquotedblright{}
in the legend for SoftDrop groomed jets). The groomed parton-level
curve is sometimes higher than its ungroomed counterpart because we
drop jets with $z_{\text{g}}<0.1$, which are also often low mass
jets.\label{fig:overallpTscan}}
\end{figure*}
The fully corrected jet mass and groomed jet mass distributions are
shown in Fig.~\ref{fig:overallpTscan} for $R=0.4$ jets with $20<p_{\text{T,jet}}<\unit[25]{GeV}/c$
in the left panel, $25<p_{\text{T,jet}}<\unit[30]{GeV}/c$ in the
middle panel, and $30<p_{\text{T,jet}}<\unit[40]{GeV}/c$ in the right
panel. As $p_{\text{T},\text{jet}}$ increases, we observe an increase
in the mean mass, $\left\langle M\right\rangle $, and groomed mass,
$\left\langle M_{\text{g}}\right\rangle $, as well as a broadening
of the distribution: jet mass and groomed jet mass mean (RMS) increase
by $25.5\%\pm0.2\%\text{(stat.)}\pm2.0\%\text{(syst.)}$ ($18.1\%\pm0.4\%\pm4.1\%$)
and $15.5\%\pm0.3\%\pm2.4\%$ ($23.2\%\pm0.4\%\pm3.6\%$) respectively,
from lowest to highest $p_{\text{T},\text{jet}}$. This is due to
the increase in the available phase space for radiation with higher
jet momentum, as expected from pQCD, although the effect is slightly
mitigated by the relative increase, as $p_{\text{T,jet}}$ increases,
in the number of quark-initiated jets, which have lower $\left\langle M\right\rangle $
than gluon-initiated jets \citep{QvG}.

We compare our fully corrected results with leading-order MC event
generators PYTHIA-8 (including a curve with hadronization turned off,
which we denote ``parton jets''), HERWIG-7, and PYTHIA-6. The latter
uses the Perugia 2012 tune \citep{Perugia} with additional tuning
to the RHIC environment \citep{STARtune}. The former two use tunes
developed for LHC kinematics - in PYTHIA-8 the Monash 2013 tune \citep{Monash},
and in HERWIG-7 the EE4C underlying event tune \citep{EE4C}. Relevant
model differences between PYTHIA and HERWIG for a jet substructure
analysis lie in the shower and hadronization mechanisms, where PYTHIA
uses a $p_{\text{T}}$-ordered shower and string fragmentation, while
HERWIG uses an angular-ordered shower and cluster hadronization.

\begin{figure*}
\begin{centering}
\includegraphics[width=0.95\textwidth]{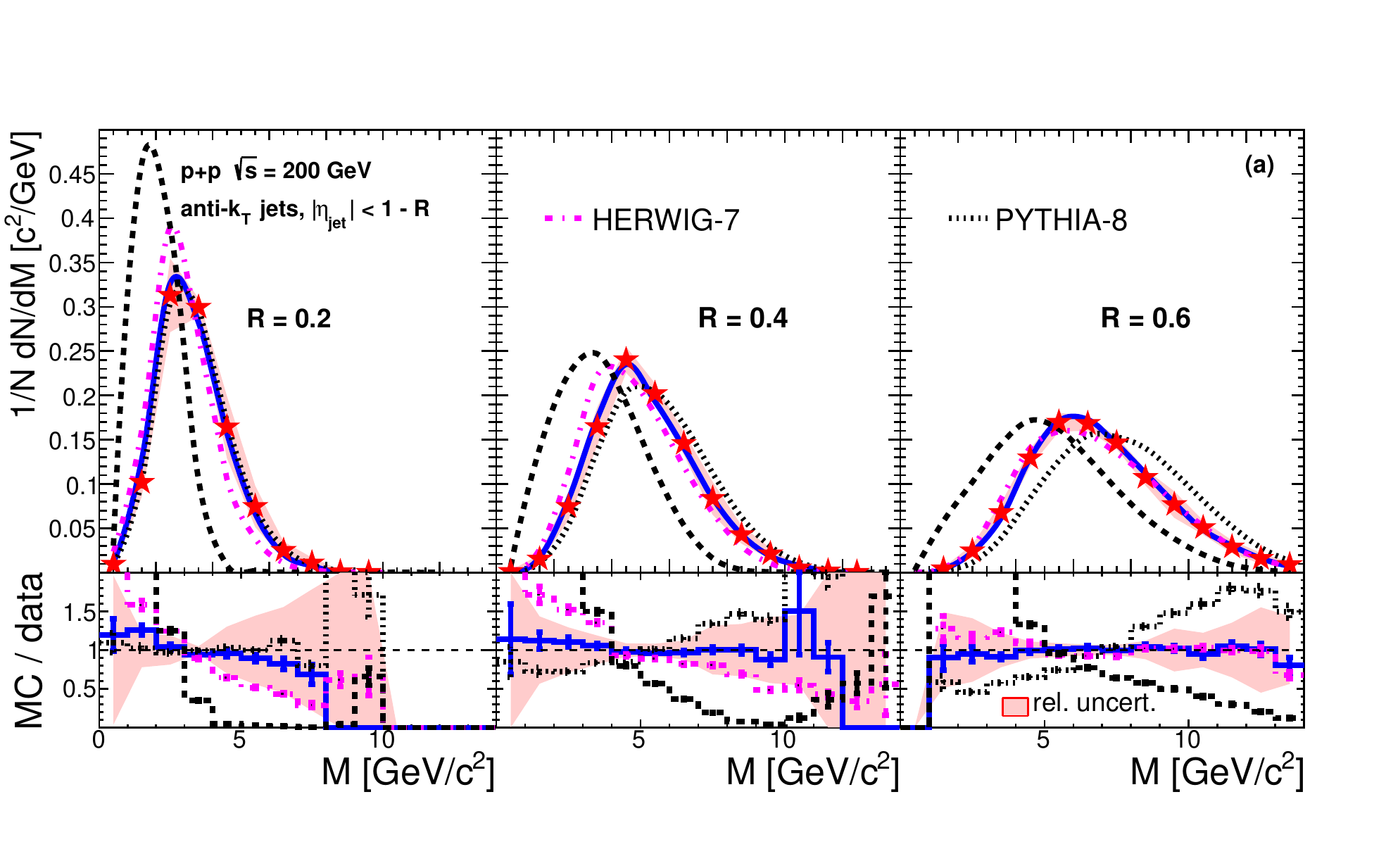}\vspace{-27bp}
\par\end{centering}
\begin{centering}
\includegraphics[width=0.95\textwidth]{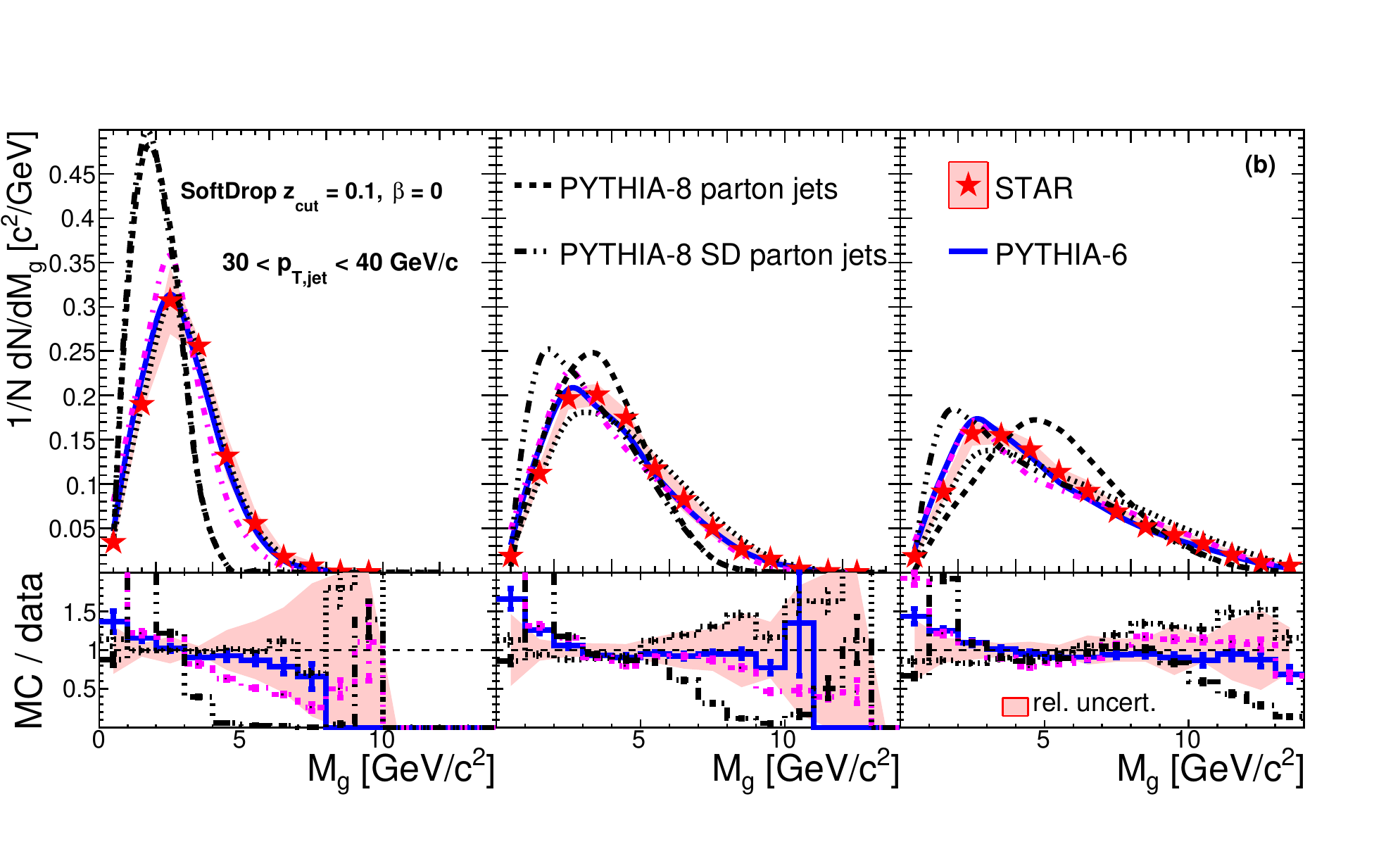}
\par\end{centering}
\caption{Distributions of (a) $M$ and (b) $M_{\text{g}}$ for anti-$k_{\text{T}}$
jets in $pp$ collisions at $\sqrt{s}=\unit[200]{GeV}$ for a single
$p_{\text{T},\text{jet}}$ selection ($30<p_{\text{T,jet}}<\unit[40]{GeV}/c$)
and varying jet radii (from left to right, $R=0.2,0.4,0.6$). The
style of the curves is the same as in Fig.~\ref{fig:overallpTscan}.
Statistical uncertainties are smaller than the size of the markers.
In (b), we additionally compare to groomed PYTHIA-8 parton-level (dot-dashed
black line, denoted \textquotedblleft SD\textquotedblright{} in the
legend for SoftDrop groomed jets) and ungroomed PYTHIA-8 parton-level
(dashed black line) to demonstrate that grooming removes NP physics
such as hadronization.\label{fig:overallRscan}}
\end{figure*}
We note that PYTHIA-6 describes the data well, within systematic
uncertainties, whereas the HERWIG-7 and PYTHIA-8 prefer lower and
higher mass jets, respectively, as observed in the slope of the ratio
of MC to data. The behavior of HERWIG-7 and PYTHIA-8 is similar to
that of HERWIG and PYTHIA (albeit different versions and tunes) in
the LHC measurement of Ref.~\citep{ATLAS1}, although with HERWIG
preferring larger mass jets and PYTHIA preferring slightly smaller
mass jets than data. We also observe from the comparison of the parton-level
to hadron-level curves from PYTHIA-8 that hadronization increases
the jet mass. In addition to the scan over $p_{\text{T},\text{jet}}$,
we show in Fig.~\ref{fig:overallRscan} the jet mass and groomed
jet mass distributions for jet radius $R=0.2$, $0.4$, and $0.6$,
for a fixed $p_{\text{T,jet}}$ from 30 to 40 GeV$/c$. We observe
that the mass increases with increasing $R$, which is expected as
the jet will encompass more wide-angle NP physics and will more often
be a gluon-initiated jet which has higher mass on average than a quark-initiated
jet.

As for the groomed mass, it exhibits similar trends of increasing
mass with increasing $p_{\text{T},\text{jet}}$ and $R$. We note,
however, that the position of the Sudakov peak is shifted up to $55\%\pm22\%\text{(stat.)}$
lower, for the largest jet radius and $p_{\text{T,jet}}$ selection,
for the groomed mass compared to the ungroomed mass distributions.
This implies that jet constituents arising from wide-angle radiation
are indeed suppressed. We also observe that the grooming procedure
results in a groomed mass that is closer to the ungroomed parton-level
curves, signaling that NP effects such as hadronization have been
reduced. Both of these effects are more significant for $R=0.6$ jets.
Finally, we note that for ungroomed mass, PYTHIA-6 is consistent with
the data in all cases, while for groomed mass, PYTHIA-6 slightly underpredicts
data for all jet radii shown. HERWIG-7 and PYTHIA-8 prefer lower and
higher mass jets, respectively, in most $R$ and $p_{\text{T,jet}}$
selections with the exception of $R=0.2$ for PYTHIA-8, and $30<p_{\text{T,jet}}<\unit[40]{GeV/}c$,
$R=0.6$ for ungroomed HERWIG-7, where they do describe the data.
The preference of HERWIG-7 and PYTHIA-8 for lower and higher $\left\langle M_{\text{g}}\right\rangle $,
respectively, is consistent with the observed preference for narrower
and wider splits, respectively, in Ref.~\citep{RaghavArXiv}.

\begin{figure*}
\begin{centering}
\includegraphics[width=1\textwidth]{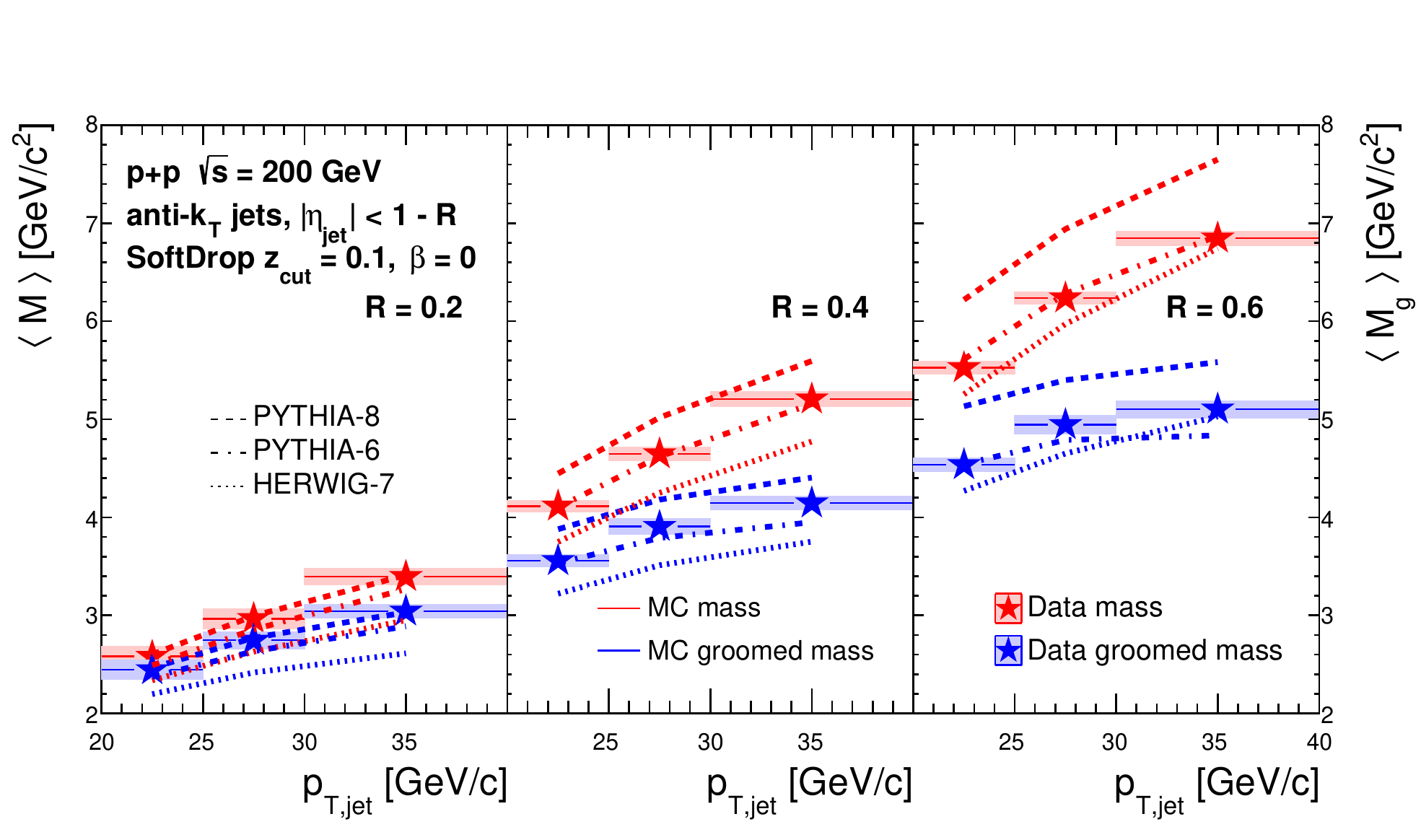}
\par\end{centering}
\caption{(Color online) Mean jet mass of anti-$k_{\text{T}}$ jets in $pp$
collisions at $\sqrt{s}=\unit[200]{GeV}$ as a function of $p_{\text{T},\text{jet}}$
($x$-axis) and jet radius, $R$ (panels). The fully corrected data
are shown as solid star markers with accompanying shaded band denoting
systematic uncertainties on the mean. The mean jet mass, $\left\langle M\right\rangle $,
is shown in red; mean groomed jet mass, $\left\langle M_{\text{g}}\right\rangle $,
is shown in blue. Statistical uncertainties are smaller than the size
of the markers. PYTHIA-6 (Perugia 2012 STAR Tune, dot-dashed line),
PYTHIA-8 (Monash 2013 Tune, dashed line), and HERWIG-7 (EE4C Tune,
dotted line) are also presented, with red denoting $\left\langle M\right\rangle $
and blue denoting $\left\langle M_{\text{g}}\right\rangle $.\label{fig:mean_1x3}}
\end{figure*}
The dependence of the mass on jet radius and $p_{\text{T},\text{jet}}$
is made more apparent by Fig.~\ref{fig:mean_1x3} in which we plot
the mean (groomed) mass for the same jet radius and $p_{\text{T,jet}}$
selections as above. Note the decrease in $\left\langle M\right\rangle $
from grooming due to removal of wide-angle radiation. Additionally,
there is an increase in the difference between the groomed and ungroomed
jet mass for higher-$p_{\text{T}}$ jets, presumably due to the fact
that high-$p_{\text{T}}$ jets are more collimated, so more of the
widest-angle radiation is captured in the jet cone before grooming,
which the grooming procedure then reduces. We also observe that for
small radius jets, the jet mass is relatively unaffected by grooming,
as much of the wide-angle radiation will already have left the jet
cone, while for large-radius jets grooming dramatically reduces $\left\langle M\right\rangle $.
There is also a much stronger dependence of $\left\langle M_{\left(\text{g}\right)}\right\rangle $
on the jet radius than on the $p_{\text{T},\text{jet}}$. The stronger
dependence on the radius is expected from the generic form of the
jet mass: $M^{2}\sim p_{\text{T},\text{jet}}\sum_{i\in J}p_{\text{T},i}\Delta R_{iJ}^{2}$,
where $p_{\text{T},i}$ is the transverse momentum with respect to
the jet axis and $\Delta R_{iJ}^{2}=\left(\Delta\eta_{iJ}\right)^{2}+\left(\Delta\phi_{iJ}\right)^{2}$
is the distance of the constituent $i$ from the jet axis \citep{Theory2}.
Lastly, we observe that in general, HERWIG-7 (PYTHIA-8) predicts a
smaller (larger) $\left\langle M_{\left(\text{g}\right)}\right\rangle $
than data, while PYTHIA-6 is roughly consistent with data, as mentioned
in the previous two paragraphs.

\begin{figure*}
\begin{centering}
\includegraphics[width=1\textwidth]{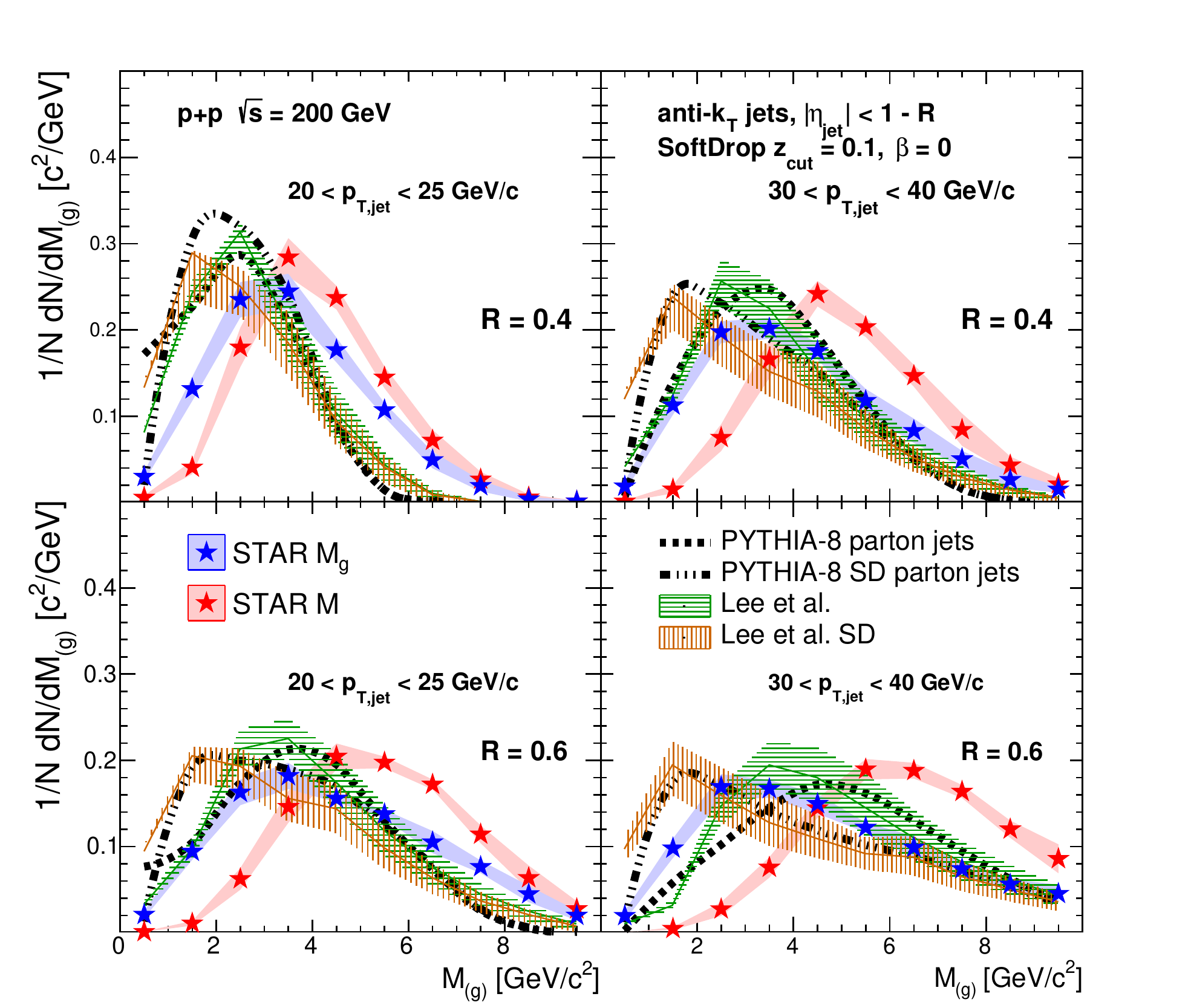}
\par\end{centering}
\caption{Comparison between hadron-level fully corrected STAR data (star markers
with shaded bands denoting systematic uncertainty) and a calculation
at next-to-leading-log accuracy at parton-level (shaded bands, which
denote QCD scale variation uncertainty) as well as parton-level MC,
for $M$ and $M_{\text{g}}$ of anti-$k_{\text{T}}$ jets in $pp$
collisions at $\sqrt{s}=\unit[200]{GeV}$ for two $p_{\text{T},\text{jet}}$
selections (from left to right, $20<p_{\text{T,jet}}<\unit[25]{GeV}/c$
and $30<p_{\text{T,jet}}<\unit[40]{GeV}/c$) and varying jet radii
(from top to bottom, $R=0.4,0.6$). The ungroomed (groomed) jet mass
data are in red (blue), while the ungroomed (groomed) jet mass calculation
is in green (orange). Statistical uncertainties are smaller than the
size of the markers. We additionally compare to groomed PYTHIA-8 parton-level
(dot-dashed black line, denoted \textquotedblleft SD\textquotedblright{}
in the legend for SoftDrop groomed jets) and ungroomed PYTHIA-8 parton-level
(dashed black line). \emph{N.B.} the calculation is dominated by NP
effects below $M\approx\unit[1]{GeV}/c^{2}$, so a linear extrapolation
to zero was done from the lowest value above this NP regime.\label{fig:theory}}
\end{figure*}

Finally, we compare the hadron-level fully corrected jet mass data
to a pQCD calculation at next-to-leading order with next-to-leading-logarithm
(NLL) accuracy at the parton-level in Fig.~\ref{fig:theory}. NP
contributions are expected to be large for small $R$, small $p_{\text{T,jet}}$,
and small $M$. Because of this large NP contribution to the calculation
for $R=0.2$ over the entire jet mass range, we show only $R=0.4$
and $R=0.6$ jets, and even for $R=0.4$ jets, the lowest $p_{\text{T,jet}}$
range is dominated by these NP contributions. Similarly, theoretical
calculations become unreliable at small $M$, so a simple linear extrapolation
of the ungroomed jet mass to $\left(0,0\right)$ is done in Fig.~\ref{fig:theory}
in order to provide a consistent overall normalization by which to
compare the calculations and data. The uncertainty on the calculation
is given by QCD scale variation by a factor of two about the chosen
values, while keeping the relation between the collinear and soft
scales and between the hard and jet scales fixed, as in Ref.~\citep{CalculationUncertainty}.
 Note also that the jet mass range for the comparison is $0<M<\unit[10]{GeV}/c^{2}$
so the normalization of the data in Fig.~\ref{fig:overallRscan}
and Fig.~\ref{fig:theory} is different.

Although NP  effects are minimized at large $R$ and $p_{\text{T,jet}}$,
we observe a significant discrepancy in the jet mass between the calculation
and data even for $R=0.6$, $30<p_{\text{T,jet}}<\unit[40]{GeV}/c$
ungroomed jets. Since N$^{n}$LO effects should be small at RHIC,
this indicates not a failing of the fixed-order calculation, but rather
the large effect of hadronization and other NP contributions to the
jet mass, e.g. multiple-particle interactions, in the STAR kinematic
regime. However, we do observe that for larger radius jets, the groomed
jet mass data and ungroomed jet mass calculation are comparable, indicating
the suppression of NP effects with the SoftDrop procedure. We notice
as well that the calculation is comparable with PYTHIA-8 both for
groomed and ungroomed jets, which suggests that most of this large
NP contribution is due to the hadronization of final-state particles.
These results can be used to extract an NP shape function for a hadronization
correction. and by extension will allow for further refinement of
theoretical quantities, such as the soft function, using the extracted
$R$-dependent shape function \citep{Theory8}.

\section{Summary}

This paper presents the first inclusive jet mass measurement in $pp$
collisions at $\sqrt{s}=\unit[200]{GeV}$ at RHIC. Both the ungroomed
and groomed jet mass are presented differentially in the jet radius
and transverse momentum. We observe trends of increasing mean and
width of the distributions with increased $R$ due to the inclusion
of more wide-angle radiation. We observe the same trends for increasing
$p_{\text{T,jet}}$ as well due to the increased phase space for radiation.
Both of these observations are consistent with pQCD. In addition,
grooming is shown to shift the mass distribution to smaller values
and reduce the position of the Sudakov peak, leaving the fully corrected
groomed jet mass amenable to comparison with NLL calculations.

Although the fully corrected data is well-described by the RHIC-tuned
leading-order MC event generator PYTHIA-6 Perugia 2012, it is not
as well-described by LHC-tuned PYTHIA-8 Monash 2013 and HERWIG-7 EE4C,
providing crucial input for further RHIC tunes. This measurement will
also serve as an important reference for future jet mass measurements
in heavy-ion collisions at RHIC in which modification due to energy
loss in the hot nuclear medium may be expected.

\section*{Acknowledgements}

We thank Kyle Lee and Felix Ringer for helpful discussions pertaining
to the jet mass. We thank the RHIC Operations Group and RCF at BNL,
the NERSC Center at LBNL, and the Open Science Grid consortium for
providing resources and support. This work was supported in part by
the Office of Nuclear Physics within the U.S. DOE Office of Science,
the U.S. National Science Foundation, the Ministry of Education and
Science of the Russian Federation, National Natural Science Foundation
of China, Chinese Academy of Science, the Ministry of Science and
Technology of China and the Chinese Ministry of Education, the Higher
Education Sprout Project by Ministry of Education at NCKU, the National
Research Foundation of Korea, Czech Science Foundation and Ministry
of Education, Youth and Sports of the Czech Republic, Hungarian National
Research, Development and Innovation Office, New National Excellency
Programme of the Hungarian Ministry of Human Capacities, Department
of Atomic Energy and Department of Science and Technology of the Government
of India, the National Science Centre of Poland, the Ministry of Science,
Education and Sports of the Republic of Croatia, RosAtom of Russia
and German Bundesministerium fur Bildung, Wissenschaft, Forschung
and Technologie (BMBF), Helmholtz Association, Ministry of Education,
Culture, Sports, Science, and Technology (MEXT) and Japan Society
for the Promotion of Science (JSPS).

\bibliographystyle{apsrev4-2}
\phantomsection\addcontentsline{toc}{section}{\refname}\bibliography{ppJetMass_bib}

\end{document}